# Unusual pattern formation on Si(100) due to low energy ion bombardment

Tanmoy Basu[1], Jyoti Ranjan Mohanty[2], T. Som[1]

[1] Institute of Physics, Sachivalaya Marg, Bhubaneswar -751 005, India

[2] Bhubaneswar Institute of Technology, Harapur, Bhubaneswar-752 054, India

E-mail: tsom@iopb.res.in

**Abstract**

In this paper evolution of silicon surface topography, under low energy ion bombardment, is investigated at higher oblique incident angles in the range of 63°-83°. Si(100) substrates were exposed to 500 eV argon ions. Different surface morphology evolves with increasing angle of incidence. Parallel-mode ripples are observed up to 67° which undergo a transition to perpendicular-mode ripples at 80°. However, this transition is not a sharp one but undergoes a series of unusual pattern formation at intermediate angles. Complete smoothening of silicon surface is observed at incident angles beyond 80°. The observed patterns are attributed to surface confined viscous flow and sputter erosion under ion bombardment.

# 1. Introduction

In recent years, low energy ion-beam induced periodic pattern formation on large area surfaces has drawn a lot of attention [1]. This single step processing route is fast, cost-effective, and has the potential to be an alternative for the conventional lithographic techniques. Different morphologies evolve in various types of materials, viz. Metals [2], semiconductors [3-6], and insulators [7, 8]. Out of these, nano-patterned semiconductor surfaces have started finding importance for applications in photonics and nanoscale magnetism [9, 10].

It is known that formation of periodic self-organized nanostructures is due to a dynamic balance among fundamental surface kinetic processes under ion bombardment [1]. The classical linear stability theory proposed by Bradley and Harper (B-H) explains many experimental observations [11]. Starting from Sigmund's theory of sputtering [12, 13], B-H theory deals with curvature dependent sputtering and describes ripple formation along with its rotation at certain oblique angle of incidence. In a different note, Umbach *et al.* described pattern formation due to ion induced viscous flow [14]. On the other hand, Madi *et al.* described pattern formation in light of stability and instability of surfaces under ion-bombardment which is governed by mass redistribution rather than sputter erosion [15]. Thus, more efforts are needed to develop a more accurate theoretical framework for qualitative prediction of low energy ion induced pattern formation. On the experimental front, the major drawbacks have been the lack of adequate experimental data, in particular at higher oblique angles of incidence, and the reproducibility of observed patterns. These give rise to the necessity of performing systematic experiments over a large angular window to learn how to control and manipulate the surface patterns by ion irradiation.

Silicon is the most widely studied system for pattern formation [3, 6, 15, 16]. Usually, parallel-mode ripples (i.e. wave vector parallel to the projected ion beam) are formed beyond an incident angle ($\theta$) of 47° which undergo a transition to perpendicular-mode ripples (wave vector perpendicular to the ion beam) at 80° [15, 17]. Over this angular window reports are available on ripple formation up to an oblique incidence of ~70° [6, 15, 16]. However, there is hardly any systematic study on Ar-ion induced pattern formation on Si in the angular window 70°≤$\theta$≤80° which would help unfolding the underlying processes leading to a transition from parallel- to perpendicular-mode ripples.

In this work, evolution of surface topography on Si(100) substrates is investigated under low energy ion bombardment over an angular window of 63°-83°. This gives rise to evolution of a gamut of patterns among which some are quite unusual (those formed at higher oblique incident angles, viz. 70°-78°) in nature. Through these intermediate patterns parallel-mode ripples are seen to undergo a transition to perpendicular-mode ripples at an incident angle of 80°. In case of an even higher angle of incidence surface smoothening is observed. Results are attributed to ion induced viscous flow and sputter erosion under energetic ion bombardment.

## 2. Experiment

The substrates used in the experiments were cut from *p*-type Si(100) wafers (B-doped, resistivity 0.01-0.02 Ω-cm). A UHV-compatible experimental chamber (Prevac, Poland) was used which is equipped with a 5-axes sample manipulator and an electron cyclotron resonance (ECR) based broad beam, filament less ion source (Tectra GmbH, Germany). The chamber base pressure was below $5\times10^{-9}$ mbar and the working pressure was maintained at $3\times10^{-4}$ mbar by using a differential pumping unit. Argon-ion energy was chosen to be 500 eV since there is hardly any systematic study at this energy for higher oblique incident angles under consideration. The beam diameter was measured to be 3 cm while the constant current density was measured to be 0.2 µA cm$^{-2}$. We used an ion flux of 1.34×10$^{14}$ ions cm$^{-2}$ s$^{-1}$ and the experiments were performed at a fixed fluence of 5×10$^{17}$ ions cm$^{-2}$ for the entire angular window of 63º-83º. In addition, measurements were carried out to study flux- and fluence-dependent evolution of surface morphology in case of selective angles.

Fig. 1 shows the schematic diagram of the experimental geometry. It is to be noted that the sample platen was covered up with a similar *p*-type Si(100) wafer and samples (1 cm×1 cm) were mounted on top of it. This type of configuration, having a sacrificial wafer, helps to avoid recording pattern formation due to incorporation of impurities originating from the sample platen and redeposition of silicon atoms originating from the sacrificial wafer.

The surface morphology was studied by *ex-situ* atomic force microscopy (AFM) in tapping mode$^{TM}$ (Asylum Research, USA). Silicon probes were used having diameter ~10 nm. For each sample large numbers of AFM micrographs were collected from different regions to check the uniformity of the pattern morphology. Root mean square roughness, *w*, was calculated from all AFM images. In order to obtain spatial correlation and periodicity two-dimensional fast Fourier transformations (2D FFT) was plotted. All these parameters were extracted from the experimental AFM images by using the WSxM software [18].

## 3. Results and discussion

Figures 2(a-h) present the AFM topographic images of silicon surface before and after exposure to argon ions at different oblique incident angles. Fig. 2(a) depicts AFM image of the pristine sample which has a smooth surface (*w*=0.09 nm). From Fig. 2(b) it is observed that ripple morphology evolves at 65º. Estimations from WSxM software show that ripples

have wavelength of 38 nm, height of 3 nm, and rms surface roughness of 1.7 nm. Careful measurements reveal that the ripple wavelength is independent of ion flux. Fig. 2(c) represents the morphology corresponding to an incident angle of 67° where parallel-mode ripples (wavelength of 60 nm, height of 3 nm, and rms roughness of 2.6 nm) are formed. In addition, occasional mounds (average dimension of 65 nm) appear on top of the ripples. Formation of such parallel-mode ripples is consistent with literature [6, 15].

Starting from this point (i.e. at 67°), where mounds on ripples are observed, unusual patterns evolve at higher incident angles which were hardly seen earlier. Fig. 2(d) presents the AFM image corresponding to an oblique incidence of 70° where mounds are observed instead of parallel-mode ripples as stated earlier. These mounds have an average dimension of 88 nm and the corresponding surface roughness is 5.7 nm. There is a mild signature that these mounds may be protruded in the direction of ion beam projection on the sample surface. For the incident angle of 72.5°, cone-like structures are observed where their apexes point towards the direction of ion beam [Fig. 2(e)]. The calculated surface roughness, mean value of the apex angle, and the base dimension are 7.8 nm, 34.8°, and 110 nm, respectively. Upon increasing the angle of incidence to 77.5° elongated structures are formed which resemble more like nano-needles. These needle-like structures (having an average length of 200 nm) lie parallel to the projection of ion beam onto the surface [Fig. 2(f)].

Further, increasing the incident angle to 80°, perpendicular-mode ripples (of wavelength 65 nm and rms surface roughness of 0.4 nm) are formed [Fig. 2(g)]. Thus, it is clear that parallel-mode ripples undergo a transition to perpendicular-mode ripples at this angle. However, this transition is not a sharp one but undergoes a series of unusual pattern formation at intermediate angles. At 82.5°, a smooth surface is observed [Fig. 2(h)] whose roughness is comparable to the pristine surface. Formation of such ultra smooth surfaces under nearly similar experimental conditions was reported earlier even at 85⁰ [17].

The insets corresponding to AFM images shown in Figs. 2(b)-(h) represent the respective 2D FFT. For instance, the FFT corresponding to the incident angle of 65⁰ depicts the obvious presence of a characteristic wavelength, which gets weakened in case of incident angle of 67⁰. On the other hand, starting from 70⁰ to 80⁰, FFTs show signature of topographical anisotropy.

From the above discussion it is observed that due to 500 eV Ar-ion bombardment many different patterns evolve over the entire angular window being considered under the present study. It may be recalled that formation of ion-induced self-organized nanostructures is generally caused due to the interplay between roughening caused due to sputter erosion and smoothening due to surface diffusion [11, 14]. However, no single theory is adequate to explain all kinds of observed experimental features. This fortifies the need to invoke possible multiple physical effects to explain our observed patterns at different incident angles.

Let us first examine the role of sputtering on the origin of pattern formation observed in our case. In order to do this, an attempt has been made to correlate the evolution of surface roughness with sputtering. Fig. 3 shows the evolution of surface roughness [obtained from AFM images shown in Figs. 2(b)-(h)] at different incident angles. We also performed TRIDYN simulation [19] to calculate sputtering yield for 500 eV Ar-ions (for the fluence of $5\times10^{17}$ ions cm$^{-2}$) corresponding to different angles. Fig. 4 shows the variation in sputtering yield with incident angle. From these figures it is seen that the roughness value peaks at 72.5⁰ whereas the sputtering yield is maximum at 70⁰. Thus, unusual pattern formation in our case may be related to sputtering.

Relief structures are observed under sputtering due to low-energy ion bombardment of single crystalline and amorphous solids [20, 21]. In the latter report, the starting point of relief structures was attributed to a mechanism proposed by Sigmund [13]. In our case, at low fluences, small relief structures are observed (images not shown). As time evolves, coarsening of these smaller relief structures starts [as shown in Fig. 2(e)] which is well described by Hauffe [21]. Fig. 5(a) shows a schematic diagram of the mechanism of coarsening of such relief structures relevant to our case. Following Hauffe, $V_n \sim jS$, where $j$ is the ion density on the surface element (which also contains the reflected ions), $S$ is the sputtering yield, and $V_n$ is the displacement velocity of a surface element in the direction of its normal; it follows that $V_a > V_b$ as $l_a > l_b$. This corroborates well with cross-sectional line profile [Fig. 5(b)] obtained from the AFM image [shown in Fig. 2(e)] corresponding to an oblique incident angle of 72.5°.

The elongated needle-like structures seen at 77.5° can be attributed to shadowing effect which is expected to cause an increase in erosion of surface protrusions compared to depressions [5, 17]. The same mechanism should be operative at 80° where the surface evolves with a relatively low rms roughness. Such a low surface roughness may be attributed to a reduced sputtering yield at 80° (Fig. 4). For an even higher glancing angle of incidence (viz. 82.5°), the observed smoothening of the surface can be explained in terms of extremely low sputtering yield where most of the ions would get reflected rather than contributing to sputtering [17].

From the above discussion it is clear that in our case sputtering has a definite role to play albeit it may not be sufficient to explain formation and behavior of all individual patterns observed at different incident angles. For instance, sputter erosion based B-H theory predicts that the observed ripple wave length should have a flux dependence and undergo a transition from parallel- to perpendicular-mode ripples at an incident angle lower than the one where the sputtering yield reaches its maximum value [11]. A careful analysis of the present experimental data (as described above) reveals that none of these two criteria is fulfilled so far rippled morphology is concerned and thus calls for invoking some other mechanism than B-H theory to explain the same. Among the other existing frame works, which describe evolution of ripple morphology, Umbach *et al.* attributed ripple formation in *a*-$SiO_2$ to surface confined viscous flow [14]. According to this model, ripple wavelength is independent of ion flux and the limit up to which surface confined viscous flow model may be applied is where the ion range is less than 5 nm [14]. Under the present conditions, both these criteria are met (SRIM-2010 simulation [22] predicts a projected range of 2.5 nm). In addition, cross-sectional transmission electron microscopy study (image not shown) reveals that the top silicon layer gets amorphous (image not shown). Thus, putting together all these facts, it would be justified to explain the evolution of present ripple morphology (up to 67°) in light of surface enhanced viscous flow model. However, the origin of mounds on the ripples (at 67°) is not clear to us.

## 4. Conclusion

In summary, evolution of surface topography has been systematically studied for *p*-Si(100) under 500 eV argon ion bombardment over an angular window of 63°-83°. Formation of parallel-mode ripples is observed up to 67° while unusual patterns are formed at several oblique incident angles in the range of 70-78°. A transition from parallel- to perpendicular-mode ripples is observed at 80° beyond which surface smoothing takes place. Thus, it is concluded that the transition from parallel- to perpendicular-mode ripples is not a sharp one but undergoes a series of unusual pattern formation (viz. mounds, cone-like, and needle-like structures) at intermediate angles. Step-by-step morphological evolution at different angles of incidence is explained in terms of surface confined viscous flow and sputter erosion. Faceted structures would be important for enhanced light absorption for photovoltaic applications. More fluence and energy dependent experiments are underway to generate substantial data for a better prediction of a pattern at any given angle.

**Acknowledgments**

We would like to acknowledge Sandeep Kumar Garg for discussion regarding various aspects of ion-beam induced pattern formation. Thanks are due to Jitendra Kumar Tripathi (Tel Aviv University, Israel) for help in TRIDYN simulation. The authors would also like to thank M. Posselt (HZDR, Germany) for providing the TRIDYN simulation code.

**Figure captions**

**Fig. 1.** (color online) Schematic diagram showing the experimental geometry.

**Fig. 2.** (color online) AFM images of Si(100): (a) Pristine and exposed to 500 eV Ar$^+$-ions to the fluence of $5\times10^{17}$ ions cm$^{-2}$ at incident angles of (b) 65º, (c) 67º, (d) 70º, (e) 72.5º, (f) 77.5º, (g) 80º, and (h) 82.5º. Corresponding height scales in (a)-(h) are 1 nm, 15 nm, 23 nm, 37 nm, 51 nm, 19 nm, 3 nm, and 2 nm, respectively. Arrows indicate projection of the ion beam on the surface. Insets show the 2D FFT obtained from the corresponding images.

**Fig. 3.** (color online) Evolution of rms surface roughness corresponding to different angle of incidence.

**Fig. 4.** (color online) Plot of sputtering yield versus angle of incidence obtained from TRIDYN simulation performed for 500 eV Ar$^+$-ions to the fluence of $5\times10^{17}$ ions cm$^{-2}$.

**Fig. 5.** (color online) (a) Schematic diagram depicting the mechanism responsible for coarsening of faceted structures at 72.5º. Arrows indicate the incident ion beam, (b) cross-sectional line profile obtained from the corresponding AFM image shown in Fig. 2(e).

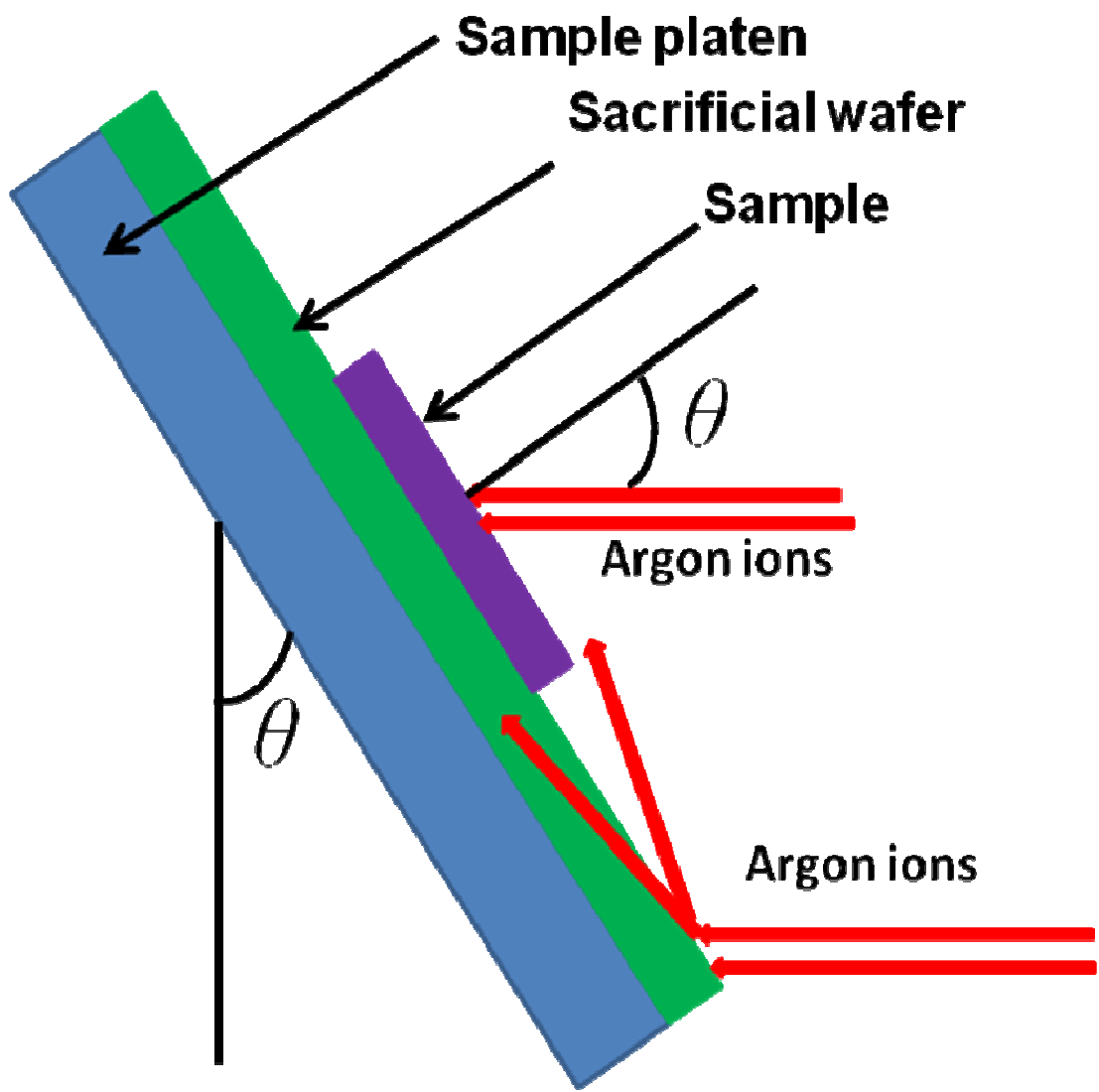

**Fig. 1**

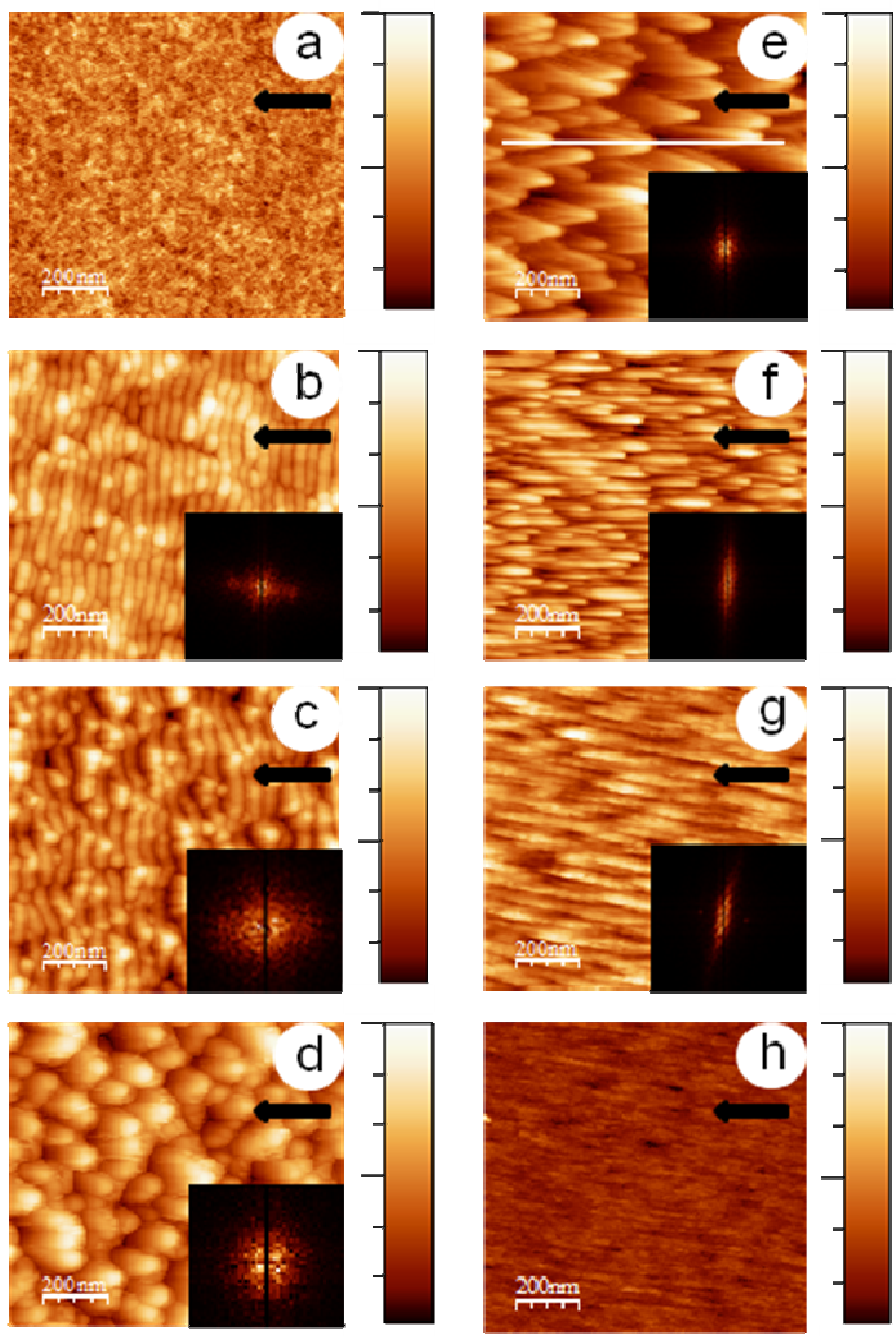

**Fig. 2**

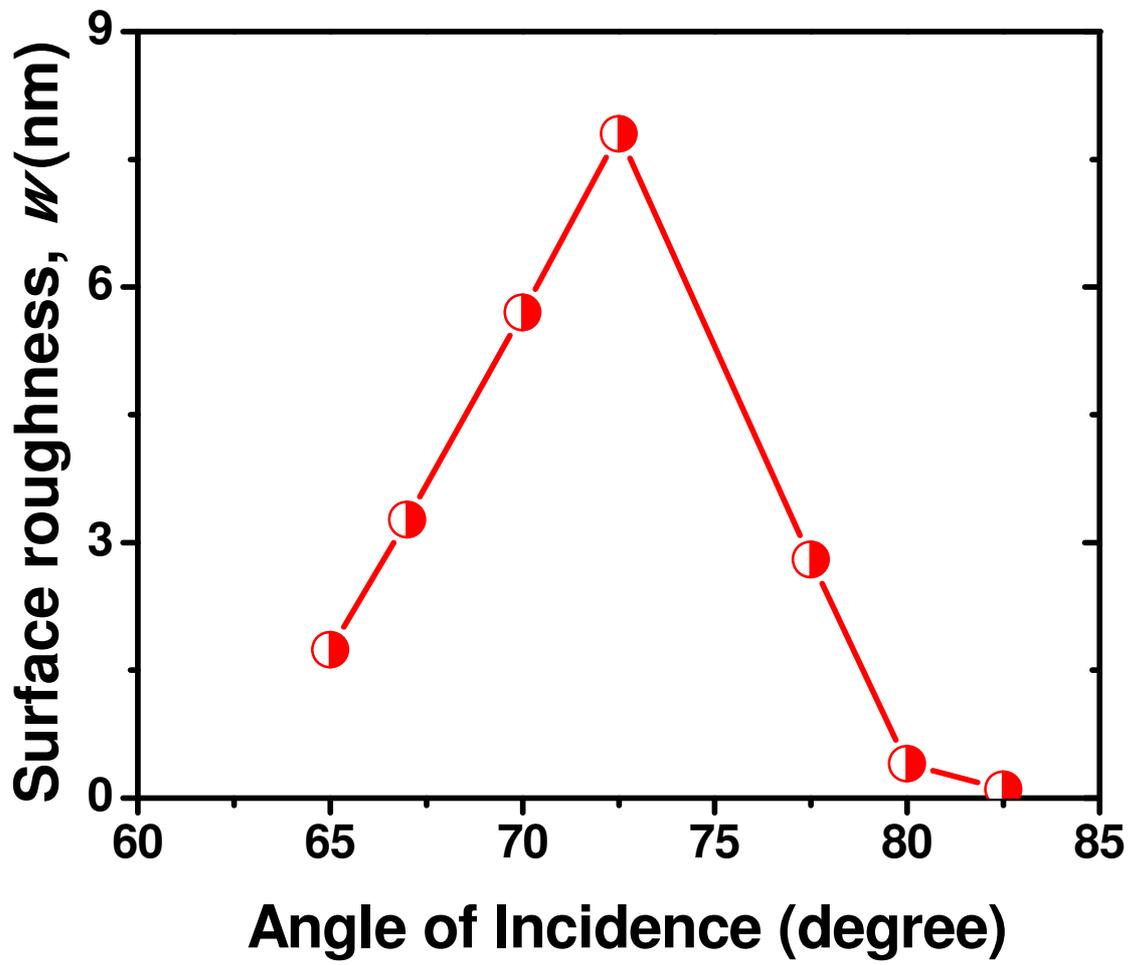

**Fig. 3**

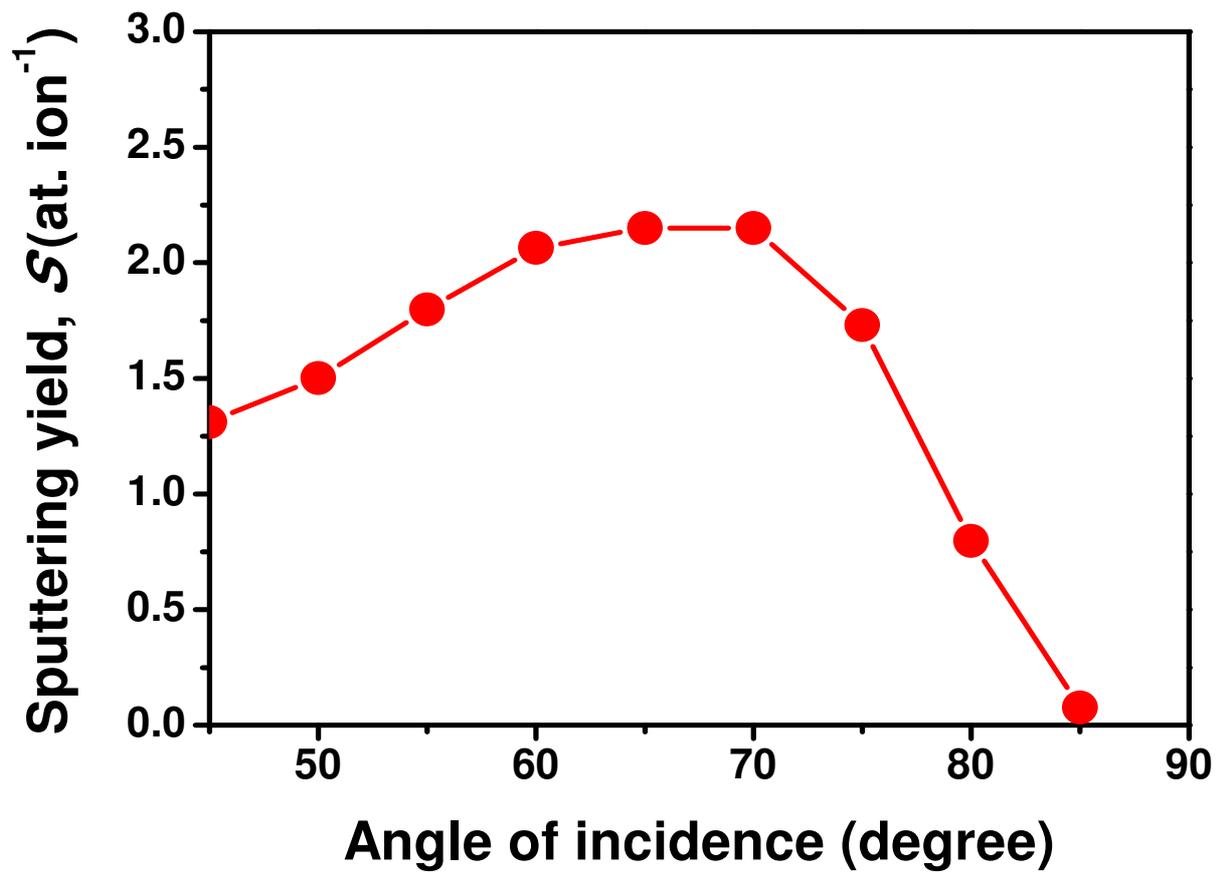

**Fig. 4**

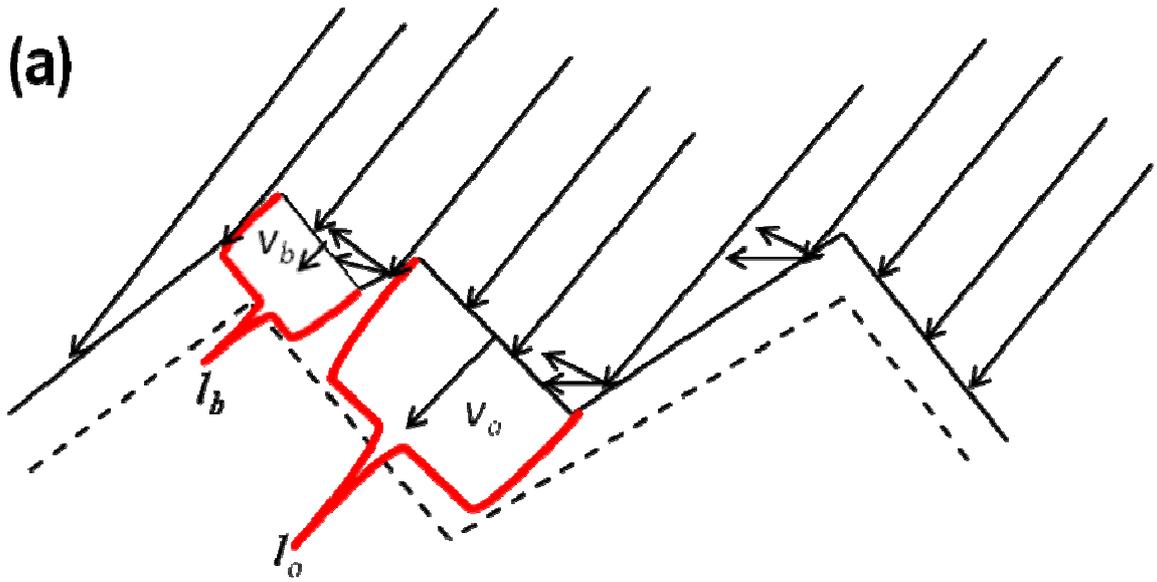
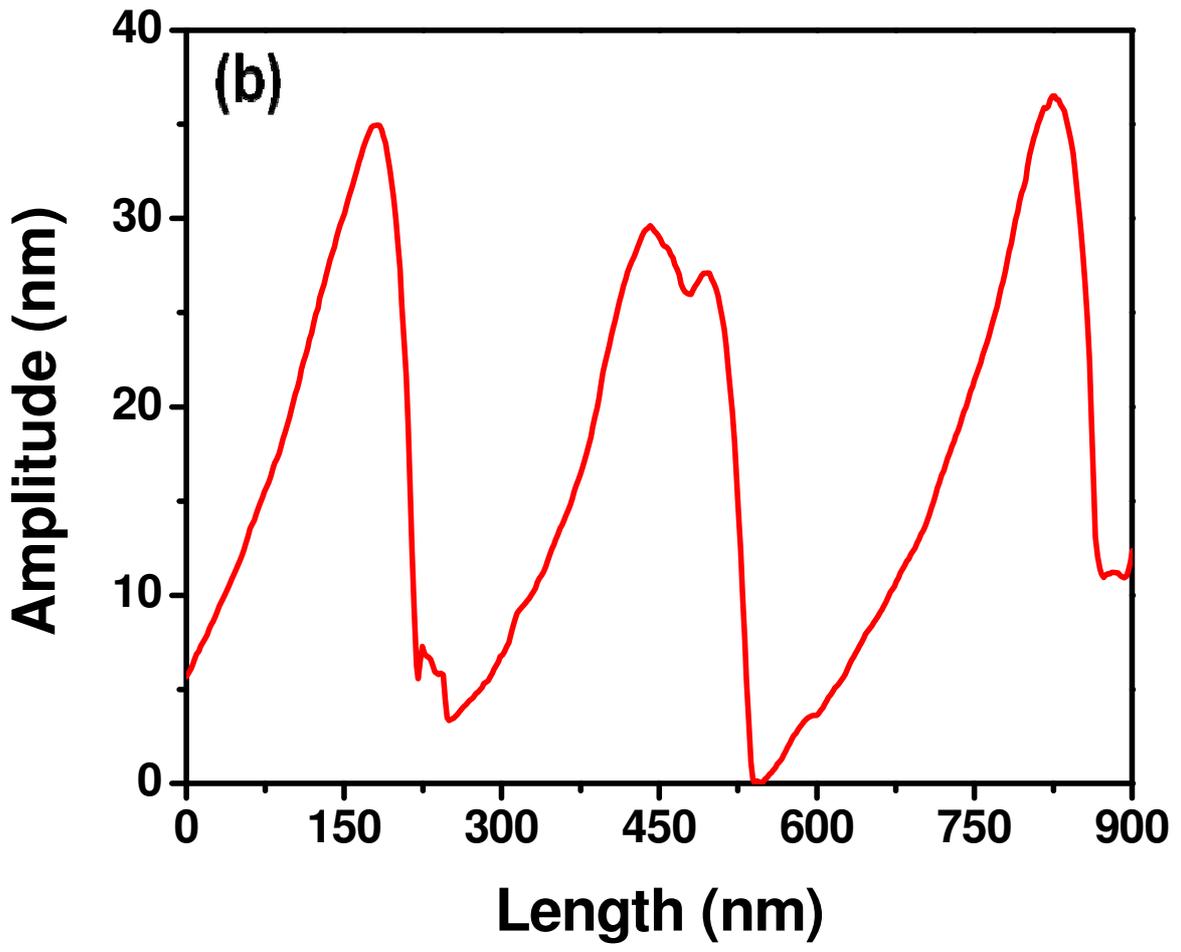

**Fig. 5**